\documentclass[preprint,aps ,nofootinbib]{revtex4}
\usepackage{graphicx}
\pdfoutput=1
\usepackage{epsfig}
\usepackage{amsmath}
\usepackage{amsfonts}
\usepackage{amssymb}
\usepackage{bm}
\usepackage{color}
\usepackage{dcolumn}
\usepackage{hyperref}
\usepackage{multirow}
\usepackage{booktabs}
\usepackage{soul}
\usepackage{ulem}
\providecommand{\U}[1]{\protect\rule{.1in}{.1in}}

\newcommand{\f}{\begin{equation}}
\newcommand{\ff}{\end{equation}}
\newcommand{\fa}{\begin{eqnarray}}
\newcommand{\ffa}{\end{eqnarray}}

\begin{document}
\title{Entanglement wedge cross section inequalities in AdS/BCFT}
\author{Pan Li $^{1,2}$}
\email{lipan@ihep.ac.cn}
\author{Yi Ling $^{1,2}$}
\email{lingy@ihep.ac.cn} \affiliation{$^1$Institute of High Energy
Physics, Chinese Academy of Sciences, Beijing 100049, China\\ $^2$
School of Physics, University of Chinese Academy of Sciences,
Beijing 100049, China }

\begin{abstract}
The entanglement wedge cross section in holographic picture provides a geometrical description of the entanglement for mixed states. In this paper we study the inequalities for the entanglement wedge cross section in AdS/BCFT duality. In the presence of the boundary in CFT, the dual entanglement wedge cross section exhibits abundant phase structures since the extremal surface may end on the brane. We present a universal treatment which is applicable for all the possible phases such that the inequalities for the entanglement wedge cross section can be proved in an algebraic manner rather than a diagrammatic manner. We show that the  entanglement wedge cross section in AdS/BCFT satisfies the same inequalities as in AdS/CFT. 

\end{abstract}

\maketitle

\section{Introduction}
Recent progress on gauge/gravity duality discloses the deep connections between spacetime geometry and quantum entanglement, and sheds light on solving the fundamental problems in black hole physics and quantum information theory. In this direction the Ryu-Takayanagi  (RT) formula \cite{Ryu:2006bv}, and the latter Hubeny-RangamaniTakayanagi (HRT) formula \cite{Hubeny:2007xt} play the central role in linking spacetime geometry to quantum entanglement. Explicitly, they conjecture that the entanglement entropy in quantum  field theory defined on boundary of AdS spacetime can be described by the extremal surface in the bulk, which is now also called holographic entanglement entropy (HEE)\cite{Ryu:2006bv,Hubeny:2007xt}. The holographic description of entanglement entropy provides an elegant geometrical realization of complicated entanglement structures, and stimulates further explorations on the relationship between the spacetime structure and quantum entanglement, such the ER=EPR conjecture\cite{Maldacena:2013xja}, the quantum chaos \cite{Shenker:2013pqa} as well as the island paradigm  recently proposed for the black hole information loss paradox\cite{Almheiri:2019psf}\cite{Penington:2019npb}.  

With the power of RT and HRT formulae, some entanglement properties such as the inequalities for entanglement entropy, which look very complicated and hard to prove from the side of quantum field theory, can be transformed into the tractable geometric problems on spacetime, which usually become more transparent and have an intuitive interpretation. For instance, the subadditivity and strong subadditivity of the entanglement entropy can be easily and intuitively proved with the language of the minimal surface in a geometric way\cite{Headrick:2007km}. Moreover, it is found that  HEE is subject to some inequalities that generic quantum states do not satisfy, such as the monogamy inequality for the mutual information (MMI)\cite{Hayden:2011ag} (at leading order of $\frac{1}{G_N}$). These inequalities give us new insight to justify which states in CFT may have classical bulk gravity duals. 
A lot of work in this subject can be found in literature, for instance see \cite{Bao:2015bfa}\cite{HernandezCuenca:2019wgh}.

Entanglement entropy is a good measure for the correlation between two subsystems $A$ and $\bar{A}$ when the whole system $(A \cup \bar{A})$ is described by a pure state $ |\phi_{A\bar{A}}\rangle$. In this case the entanglement entropy of $A$ is equal to that of $\bar{A}$: $S(A)=tr(\rho_A ln \rho_A)=S(\bar{A})=tr(\rho_{\bar{A} }ln \rho_{\bar{A}})$, where $\rho_A$ or $\rho_{\bar{A}}$ is the reduced density matrix of the subsystem $A$ or $\bar{A}$, respectively. However, when the union of two subregions ($A \cup B\equiv AB$) is not described by a pure state, for instance in a tripartite system with three subregions $A, B$ and $C$, then the correlations between $A$ and $B$ can not be precisely captured by either the entanglement entropy $S(A)$ or $S(B)$, and in general $S(A)$ is not equal to $S(B)$ anymore. As a matter of fact, one has to face this problem if one intends to understand the fine structure of correlations in an entangled many-body system. In this circumstance, one traditional way is to introduce the notion of mutual information, which is defined as $I(A:B)=S(A)+S(B)-S(AB)$, 
to describe the correlations between $A$ and $B$. The other way is to purify the subsystem $A B$ by embedding it into a larger system which is described by a pure state, and then one finds the minimum of the entanglement entropy between two subsystems including $A$ and $B$ respectively, among all the possible purifications, which is called the entanglement of purification ($E_P$). In a word, the entanglement of purification $E_P$ is proposed to measure the correlations between $A$ and $B$ in mixed states.
Nevertheless, the above definition of $E_P$ involves in an extremal process, and the practical operation is always cumbersome and complicated and thus evaluating $E_P$ is a notoriously difficult task in field theory\cite{Hirai:2018jwy,Caputa:2018xuf}. Thanks to holographic duality, inspired by the geometric description of entanglement entropy, it is conjectured that the entanglement of purification $E_P$ may also be evaluated by a geometric quantity, namely the entanglement wedge cross section ($E_W$)\cite{Takayanagi:2017knl}\cite{Nguyen:2017yqw}. Given two disjoint regions A and B on the boundary, one can firstly obtain the minimal surface ($\Gamma_{AB}^{min}$) of $AB$ as the dual of entanglement entropy. Then the bulk region surrounded by AB and $\Gamma_{AB}^{min}$ is the entanglement wedge $M_{AB}$. $E_W$ is defined as the area of the minimal surface which separates $M_{AB}$ into two bulk subregions  associated to A and B, respectively. Some evidences
supporting this conjecture can be found in \cite{Hirai:2018jwy}\cite{Caputa:2018xuf}, and further investigations on $E_W$ can be found in \cite{Umemoto:2018jpc}-\cite{Liu:2019qje}. Alternatively, there are some other quantities proposed in field theory as the avatars for $E_W$, such as reflected entropy\cite{Dutta:2019gen},  entanglement negativity\cite{Kudler-Flam:2018qjo} and odd entanglement entropy\cite{Tamaoka:2018ned}, which exhibit the similar properties with $E_W$. 

The conjecture of ``$E_W=E_P$'' was proposed partly based on the fact that both $E_W$ and $E_P$ satisfy a set of identical inequalities. Some of inequalities were previously proved for $E_W$ in the context of AdS/CFT \cite{Takayanagi:2017knl}, with the focus on some typical and specific diagrams (For instance, each CFT subregion in inequalities is connected.). In this diagrammatic proof, one need firstly determine the specific phase of the minimal surface as well as the corresponding entanglement wedge, which usually depends on the size and shape of the subregions and their relative positions.
The existence of many possible phases for the minimal surface will make the diagrammatic proof tediously complicated and even become an obstacle.
This indeed happens when boundary subregions consist of many disconnected parts, i.e.$A=A_1 \cup A_2 \cup ...$, and in particular when we intend to prove these inequalities in the context of   AdS/BCFT\cite{Takayanagi:2011zk}, where the presence of the boundary of CFT brings lots of new entanglement phases from the bulk point of view \cite{Chou:2020ffc}. So in this paper we intend to develop an algebraic proof for these inequalities in a quite general setup which applies to the case that the subregions of CFT may have disconnected parts, and more importantly to the case that the bulk spacetime may contain branes such as in the context of AdS/BCFT. Our strategy is to define some general sets for the candidates of the HEE and $E_W$. Then based on the structure of these sets and by virtue of ``wedge nesting'' \cite{Czech:2012bh}-\cite{Headrick:2013zda}, we provide a general treatment which is applicable for various entanglement phases so as to avoid the complicated enumerations of all possible diagrams. With the similar spirit the inequalities for HEE have been investigated in \cite{Hayden:2011ag,Bao:2015bfa,Headrick:2013zda}.

The paper is organized as follows. In next section we will construct the general setup
for the proof of $E_W$ inequalities, which largely relies on the definitions of HEE and $E_W$, as well as the property of ``wedge nesting". We will introduce a candidate set for HEE and a candidate set for $E_W$ in both contexts of AdS/CFT and AdS/BCFT correspondence.
In section 3, we will apply the definition of those sets to accomplish the proof for the inequalities of $E_W$ in an algebraic manner, which is applicable for different entanglement phases. We compare this method with the diagrammatic proof in the section of discussion.

\section{The general setup }
 We start with a general asymptotically AdS spacetime with CFT living on the boundary. In the case of AdS/BCFT\cite{Takayanagi:2011zk}, the bulk dual of the boundary conformal field theory (BCFT) contains brane Q. We will consider the simplest case where the brane $Q$ is embedded into AdS spacetime  perpendicular to the conformal boundary. We consider a static time-reversal symmetric slice in the bulk. $M$ represents the whole bulk spatial region, and $\partial M$ is the asymptotic  AdS boundary where the BCFT lives. In addition, we will prove the inequalities of entanglement in the large N limit, which implies that only the leading term is taken into account and the bulk fields satisfy the classical Einstein equation. 

 Now, consider subregions on the boundary $A$, $B$, $C$ .... It is assumed that they do not overlap with any non-zero size  (but they can be adjacent). Moreover, we stress that each of the subregion can be the union of many disconnected parts, i.e. $A=A_1 \cup A_2 \cup ..., \quad B=B_1 \cup  B_2 \cup ...$. In addition, we remark that in this paper all elements linked by the symbol $\cup$ do not overlap with one another. The entanglement wedge in the bulk associated with the subregion $X$  on the boundary is denoted as $M_X$.

\subsection{HEE}
In this subsection we will review the definition of holographic entanglement entropy  and then define the set consisting of the candidates for the minimal surface.
\subsubsection{HEE in AdS/CFT correspondence}

Given a  subregion $A$ on the boundary of AdS, $\Gamma_A$  is defined as a surface in the bulk satisfying the following conditions:
\begin{enumerate}
\item $ \Gamma_A$ have the same boundary with $A :\quad\partial A=\partial \Gamma_A$.
\item$ \Gamma_A$ is homologous with $A: \quad \exists R_A\subset M, \partial R_A=A\cup \Gamma_A$.
\end{enumerate}

We collect all these elements to define a set $\boldsymbol{\mathit{\Gamma_A}} $ for candidates of the minimal surface
\begin{equation}
\boldsymbol{\mathit{\Gamma_A}}
=\left\{1. \ \ \Gamma_A| \partial A=\partial \Gamma_A;2. \ \ \exists R_A\subset M, \partial R_A=A\cup \Gamma_A \right\}.
\end{equation} 
As proposed in \cite{Ryu:2006bv}, for Einstein gravity  the entanglement entropy in the dual field theory can be identified as \footnote{For the CFT dual to Einstein gravity we have the minimal surface prescription for the corresponding entanglement entropy, but for CFT dual to other general gravity theories such as higher derivative gravity one needs to minimize some other formulae \cite{Dong:2013qoa}\cite{Miao:2014nxa} to obtain the right entanglement entropy. For simplicity, in this paper we restrict our focus on the Einstein gravity only such that the entanglement entropy can always be represented by the corresponding minimal surface. We thank the anonymous referee for drawing our attention to the minimization prescription for higher derivative gravity.}
\begin{equation}
S(A)=\underset{\Gamma_A}{min}\frac{\mathcal{A} (\Gamma_A)}{4G_N},\label{hee}
\end{equation}

\begin{figure}
\centering 
\includegraphics[width=0.9\textwidth]{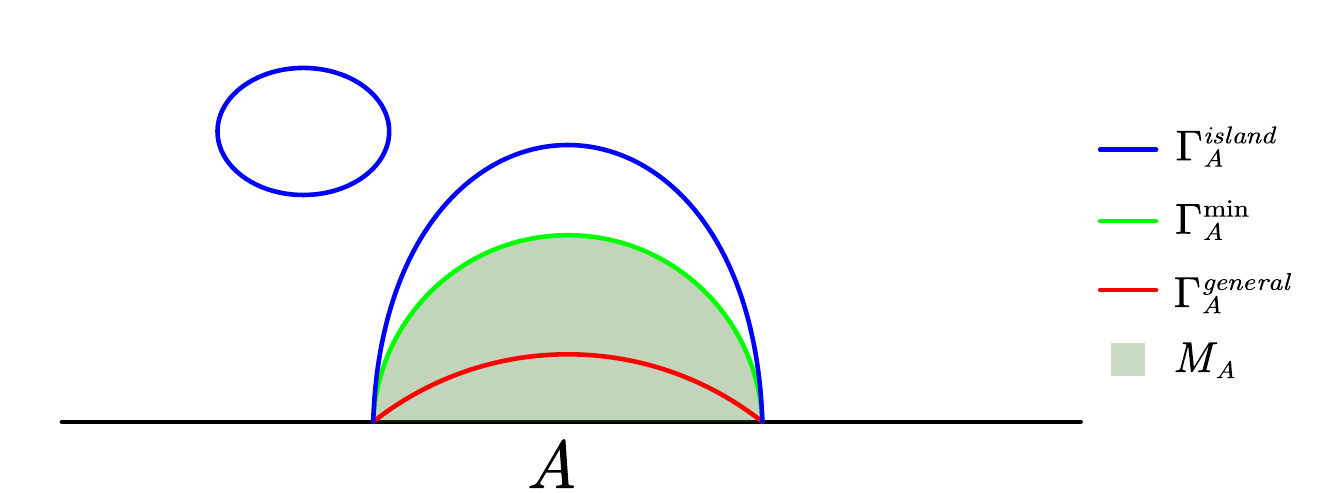}
\caption{\label{fig:candidatesofHEE} 
The schematic demonstration of three different candidate surfaces for the minimal surface, each of which has the same boundary with $A$ and homologous with $A$. We pick the one that has the minimal area to be the minimal surface, which is plotted in green. The corresponding entanglement wedge $M_A$ is shaded in grey.}
\end{figure}

where $\mathcal{A} (\Gamma_A)$ represents the area of  $\Gamma_A$. We define the element with the minimal area in set $\boldsymbol{\mathit{\Gamma_A}}$ as $\Gamma^{min}_A$, and the above formula means that the area of $\Gamma^{min}_A$  over $4G_N$ is interpreted as the entanglement entropy between the subregion $A$ and its complementary. Correspondingly, the interior region $M_A$ surrounded by $A$ and $\Gamma^{min}_A$ is  called the entanglement wedge of A (at a static slice). In Fig.(\ref{fig:candidatesofHEE}) we plot three typical configurations for candidates of the minimal surface.

\subsubsection{HEE in AdS/BCFT correspondence}
\begin{figure}
\centering 
\includegraphics[width=1\textwidth]{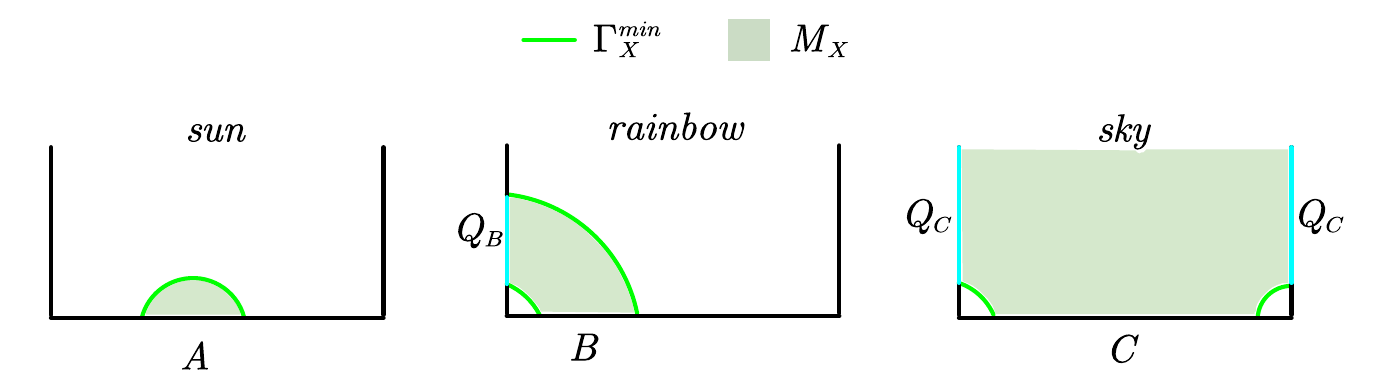}
\caption{\label{fig:bcftstructure} 
Three possible phases for the minimal surface and the corresponding entanglement wedge when the subregion is located at different positions relative to the brane $Q$ in boundary CFT \cite{Chou:2020ffc}.
}\end{figure}

The boundary of CFT may be implemented by embedding branes into the bulk spacetime. We consider the simplest case where branes are set to be perpendicular to the boundary of AdS such that the backreaction of the branes can be ignored. In this situation, the possible configurations of the minimal surface become slightly different, which may end on the branes. Therefore, we define 
the candidate set for the  minimal surface for BCFT as\cite{Miao:2017gyt}\cite{Chang:2018pnb}:
\begin{equation}
\boldsymbol{\mathit{\Gamma_A}}=\left\{\Gamma_A| 1. \ \  \partial \Gamma_A|_{\partial M}=\partial A,\ \ \partial \Gamma_A|_Q=\partial Q_A;\ \ 2. \ \  \exists R_A\subset M, \partial R_A=A\cup Q_A\cup \Gamma_A \right\},\label{heeb}
\end{equation}

and
\begin{equation}
S(A)=\underset{\Gamma_A}{min}\frac{\mathcal{A} (\Gamma_A)}{G_N},
\end{equation}
where $Q$ denotes branes in AdS, and $Q_{A}$ represents the part of the boundary of $R_A$ which locates in the brane $Q$.
It is noticed that the area of $Q_A$  is not included as part of $S(A)$. The different phases
of the minimal surface in AdS/BCFT are illustrated in Fig. \ref{fig:bcftstructure} and we name these phases following \cite{Chou:2020ffc}.

\subsection{$E_W$ in AdS/BCFT correspondence}

In this subsection, we would like to extend the notion of entanglement wedge cross section $E_W$ which originally proposed in \cite{Takayanagi:2017knl}\cite{Nguyen:2017yqw} in AdS/CFT as the holographic duality of $E_P$ to the context of AdS/BCFT \footnote{The similar extension for $E_W$ in AdS/BCFT appears in \cite{Li:2021dmf, Ling:2021vxe}, where $E_W$ is conjectured to be dual to the reflected entropy rather than the entanglement of purification, and the effects of islands and quantum corrections due to bulk fields are taken into account as well.}. One new feature of $E_W$ in AdS/BCFT duality is that $E_W$ may end on the brane Q. Here we will introduce a set for candidates of the minimal entanglement wedge cross section.\par

In comparison with the holographic description of entanglement entropy, the key difference in $E_W$ is that it always involves in at least two subregions $A$ and $B$, and we need to divide the entanglement wedge $M_{AB}$ into two parts to define the cross section, which looks more complicated. However, we remark that it is very worthy introducing such sets to collect all the candidates of $\Sigma_{A:B}^{min}$. Once such sets are constructed, then the proof of inequalities will become very concise and elegant, as we will demonstrate in next sections. For clearness, we divide the procedure of figuring out $E_W(A:B)$ into following steps.

\begin{figure}[ht]
\centering 
\includegraphics[width=1\textwidth]{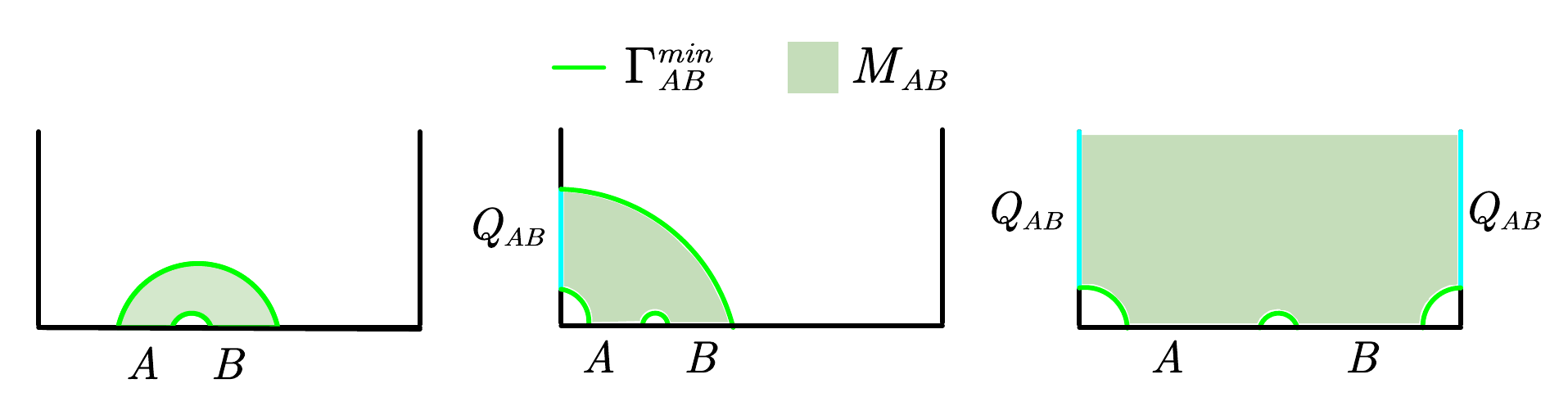}
\caption{\label{fig:bcftsabphases} 
Three typical phases of S(AB), which depend on relative positions of $A$, $B$ and the brane Q.}
\end{figure}

\begin{enumerate}
\item Given subregions $A$ and $B$ on the boundary CFT, one can directly figure out the minimal surface $\Gamma_{AB}^{min}$. Then the corresponding entanglement wedge of $AB$ in the bulk is denoted as $M_{AB}$, which is defined by\begin{equation}
    \partial M_{AB}=A\cup B \cup \Gamma_{AB}^{min}  \cup Q_{AB},
\end{equation}  as we demonstrate in Fig.(\ref{fig:bcftsabphases}).

\item Division: We divide  $\Gamma_{AB}^{min}$  and $Q_{AB}$  into two individual parts respectively \begin{align}
    \Gamma_{AB}^{min}=\Gamma_{AB}^A \cup \Gamma_{AB}^B, \\
    Q_{AB}= Q_{AB}^A \cup Q_{AB}^{B}. 
\end{align}

\item Recombination: We  define \begin{align}
    \widetilde{A}=A \cup \Gamma_{AB}^A\cup Q_{AB}^{A},\\ \widetilde{B}=B \cup \Gamma_{AB}^B\cup Q_{AB}^{B}, 
\end{align}such that the contour of the entanglement wedge $M_{AB}$ is composed of these two parts, namely  $\partial M_{AB}=\widetilde{A} \cup \widetilde{B}$.

\item  Minimization: Usually the procedure of the minimization has the following steps. Firstly, one specifies $\widetilde{A}$ and $\widetilde{B}$, and then finds the minimal surface $\Sigma^{min}_{A:B}(\widetilde{A}$) homologous to $\widetilde{A}$, as illustrated in the left plot of Fig.(\ref{fig:WABSET}).  Secondly, varying $\widetilde{A}$ and $\widetilde{B}$, one finds the corresponding $\Sigma^{min}_{A:B}(\widetilde{A}$) again, as illustrated in the right plot of Fig.(\ref{fig:WABSET}). Finally, one figures out the minimal cross section $\Sigma_{A:B}^{min}$ among $\Sigma^{min}_{A:B}(\widetilde{A}$)  for all the possible $\widetilde{A}$, namely  
\begin{equation}
E_W(A:B)=
{min}\frac{\mathcal{A}\left(\Sigma_{A:B}^{min}(\widetilde{A})
\right)}{4G_N}. 
\end{equation}

We may rephrase the above procedure of minimization by defining a set $\boldsymbol{\mathit{\Sigma^{min}_{A:B}}}$, which  collect all $\Sigma_{A:B}^{min}(\widetilde{A})$ as the candidates for the
minimal entanglement wedge cross section
\begin{equation}\begin{split}
    \boldsymbol{\mathit{\Sigma^{min}_{A:B}}}= \{\Sigma_{A:B}^{min}(\widetilde{A}) | \ \ &1. \ \partial \Sigma _{A:B} =\partial\widetilde{A}; \quad  2.\ \exists R_A\subset M_{AB},  \partial R_A=\Sigma_{A:B}\cup\widetilde{A};\\ & 3.\ \ \Sigma_{A:B}^{min}(\widetilde{A}) \text{ is the minimal surface among }\Sigma_{A:B} \text{ with }  \widetilde{A} \text{ specified} \}.    
\end{split}
\end{equation}

\item Extension: for our purpose we may remove condition 3 in $ \boldsymbol{\mathit{\Sigma^{min}_{A:B}}}$  to define a larger set
\begin{equation}
\boldsymbol{\mathit{\Sigma_{A:B}}}= \{\Sigma_{A:B}| 1. \ \  \partial \Sigma _{A:B} =\partial\widetilde{A};  \quad 2. \ \ \exists R_A\subset M_{AB}, \partial R_A=\Sigma_{A:B}\cup\widetilde{A}\}.\label{es}
\end{equation}

\begin{figure}
\centering 
\includegraphics[width=0.78\textwidth]{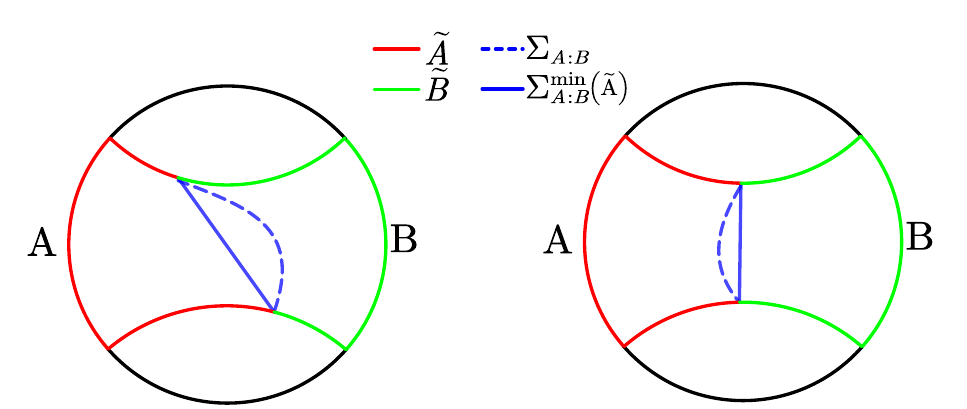}
\caption{\label{fig:WABSET} 
The process demonstration of figuring out the minimal cross section $\Sigma_{A:B}^{min}$ for CFT, which is identified as the one with the minimal area among $\Sigma^{min}_{A:B}(\widetilde{A})$ for all the possible $\widetilde{A}$.}
\end{figure}
  
\begin{figure}[ht]
\centering 
\includegraphics[width=1\textwidth]{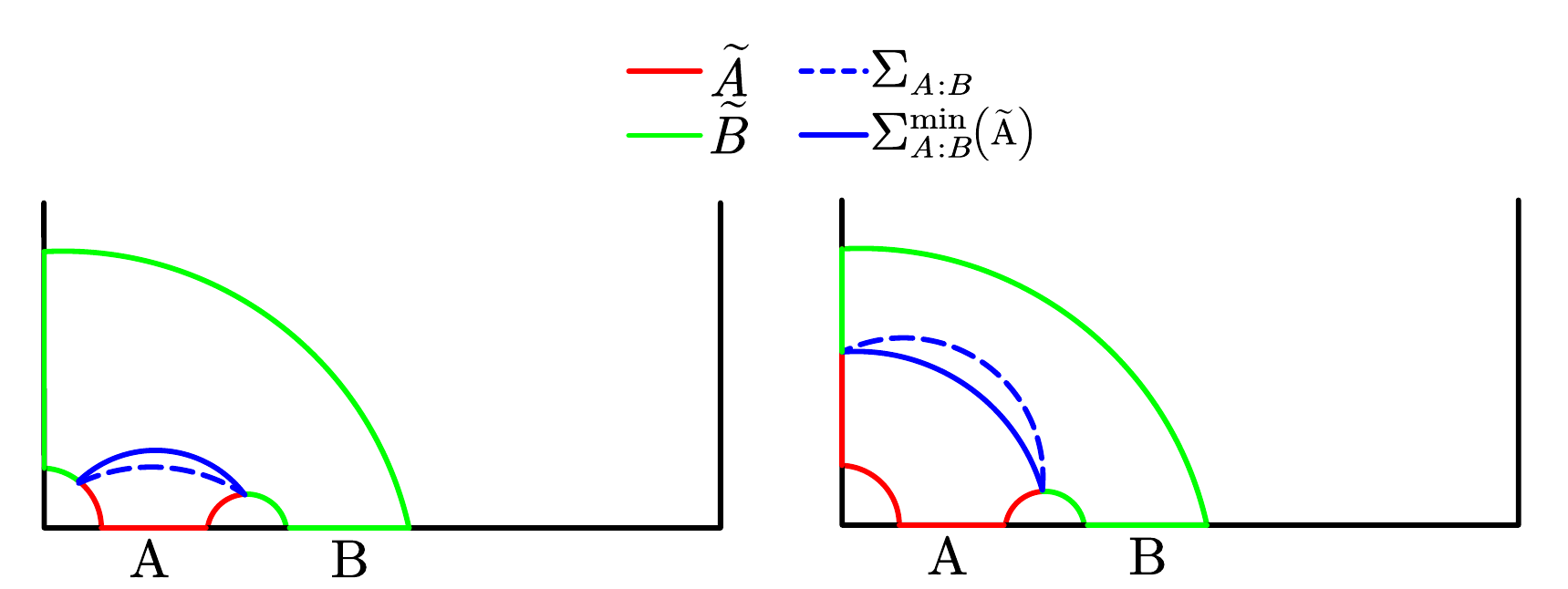}
\caption{\label{fig:WABSETBCFT} 
The process demonstration of figuring out the minimal cross section $\Sigma^{min}_{A:B}$ for BCFT, where there are more possible phases for $\Sigma_{A:B}^{min}(\widetilde{A})$. We just show two typical phases here.
}\end{figure} 

In fact, after the extension, all surfaces homologous to all possible $\widetilde{A} $ are collected in this set (In contrast, $\boldsymbol{\mathit{\Sigma^{min}_{A:B}}}$ just collects the cross sections with the minimal area which are homologous to any possible $\widetilde{A} $), as illustrated in Fig.(\ref{fig:WABSET}) and (\ref{fig:WABSETBCFT}). Roughly speaking, we could understand this set as the collection of all the possible cross sections which divide the entanglement wedge $M_{AB}$ into two parts and successfully separate $A$ and $B$. Moreover, by definition  we know that the element with the smallest area in $ \boldsymbol{\mathit{\Sigma^{min}_{A:B}}}$ must be the element with the smallest area in $ \boldsymbol{\mathit{\Sigma_{A:B}}}$ as well. Namely

\begin{equation}E_W(A:B)=
{min}\frac{\mathcal{A} \left(\Sigma_{A:B}\right)}{4G_N}.\label{EW}
\end{equation}
 
 From now on, we restrict  $\Sigma_{A:B}^{min}$ to represent the minimal-area element in $ \boldsymbol{\mathit{\Sigma_{A:B}}}$. In other words, $\mathcal{A} \left(\Sigma_{A:B}^{min}\right)=E_W(A:B)$, where for convenience, we have also simply set $4G_N=1$. 
 \subsection{Wedge nesting}
To prove some inequalities of $E_W$ in an algebraic way, it is essential to employ the properties called ``wedge nesting'', which was investigated in \cite{Czech:2012bh,Wall:2012uf,Headrick:2014cta,Headrick:2013zda}. Basically, ``wedge nesting''  contains the following two statements about the relation between boundary regions and the corresponding entanglement wedge
\begin{enumerate}
\item 
 $A \cap B=\emptyset \Rightarrow $ $M_A \cap M_B=\emptyset$.
\item  $A \subset B \Rightarrow M_A \subset M_B$.
\end{enumerate}
These two properties play a key role in the proof of some inequalities for $E_W$, as we show in the next section.

\end{enumerate}\par

\section{The inequalities of  $E_W$ in AdS/BCFT duality}

Now with all the ingredients at hand, we intend to prove the following five inequalities of  $E_W$  in the context of AdS/BCFT by geometric methods. The expressions such as $S(A)$ and $I(A:B)$ are understood as their geometric avatars, which are evaluated by the area of the corresponding minimal surfaces.
\begin{gather}
\min  \left \{ S(A),S(B) \right \} \ge E_W(A:B), \\
E_W(A:B) \ge \frac{1}{2} I(A:B), \label{inequality 2}\\
E_W(A:BC) \ge  E_W(A:B), \label{inequality 3}\\
E_W(A:BC) \ge \frac{1}{2}I(A:B)+\frac {1}{2}I(A:C),\\
E_W(AB:CD) \ge E_W(A:C)+E_W(B:D).
\end{gather}\par

Previously such inequalities have been considered for AdS/CFT  in \cite{Takayanagi:2017knl}. In addition, some inequalities for holographic entanglement entropy in the context of AdS/BCFT was presented in\cite{Chou:2020ffc}.

\begin{figure}[ht]
\centering 
\includegraphics[width=1\textwidth]{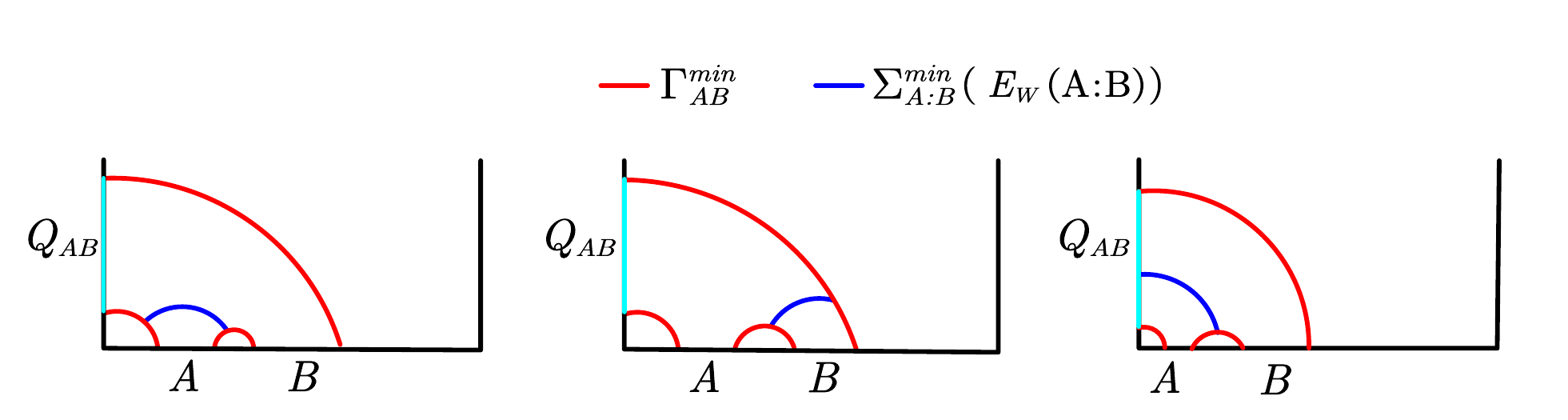}
\caption{\label{fig:bcftewcsdifferentphases} 
The phase transition of $E_W(A:B)$  when changing the length of A, B and their relative positions with brane Q.}
\end{figure}

In principle, we can prove the above inequalities as the way performed in  \cite{Chou:2020ffc}. Classifying and listing all possible entanglement phases appearing in those inequalities, visualizing each phase with an individual diagram, and then showing the proof in all possible diagrams. But as we mentioned before, the boundary effect of BCFT brings many new entanglement phases in the dual bulk. Even just for the proof of the inequalities for HEE in BCFT, there are a large number of diagrams to show \cite{Chou:2020ffc}, not to mention the phases of entanglement wedge cross sections which are constructed  based on the phases of entanglement entropy, and thus become more complicated. (seeing Fig.(\ref{fig:bcftewcsdifferentphases}) for an example.) Therefore, with the help of the candidate sets $\boldsymbol{\mathit{\Gamma_A}}$,  $\boldsymbol{\mathit{\Sigma_{A:B}}} $, 
here we develop an algebraic way to prove the inequalities of $E_W$, which does not depend on the entanglement phases.  \par

The general idea is following: the geometric correspondence of entropy inequalities in holography are essentially about the area of different surfaces. It is noticed that all inequalities we intend to prove can be translated into the following form: $\mathcal{A}(X) \ge \mathcal{A}(\Gamma^{min}_A) $ or $\mathcal{A}(Y) \ge \mathcal{A}(\Sigma^{min}_{B:C}) $. (A, B and C are just arbitrarily chosen.) Recall that  $\Gamma^{min}_A $ and $\Sigma^{min}_{B:C}$ are the minimal surfaces in $\boldsymbol{\mathit{\Gamma_A}}$ or $\boldsymbol{\mathit{\Sigma_{B:C}}}$.
If one can prove that the surface $X$ or $Y$ on the left-hand side belongs to the set $\boldsymbol{\mathit{\Gamma_A}}$ or $\boldsymbol{\mathit{\Sigma_{B:C}}}$, then above inequalities will be satisfied, according to the  definitions of $\boldsymbol{\mathit{\Gamma_A}}$ and $\boldsymbol{\mathit{\Sigma_{B:C}}}$. In this way one great advantage is that one does not need to figure out the specific phases of surfaces $X$ or $Y$, on the contrary, one just needs to show they belong to  $\boldsymbol{\mathit{\Gamma_A}}$ or $\boldsymbol{\mathit{\Sigma_{B:C}}}$, respectively. 

\subsection{$\min  \left \{ S(A),S(B) \right \} \ge E_W(A:B) $}

This inequality indicates that both holographic entanglement entropy $S(A)$ and $S(B)$ must be larger than or equal to the entanglement wedge cross section between subregion $A$ and $B$. It can be proved as follows:\par
Obviously, by definition, the quantity $E_W(A:B)$ on the right-hand side in this inequality
corresponds the surface $\Sigma_{A:B}^{min}$ (the minimal-area element in $\boldsymbol{\mathit{\Sigma_{A:B}}}$), 
 therefore following the above strategy we just need to prove that the surface associated with any quantity on the left-hand side is an element belonging to the set $\boldsymbol{\mathit{\Sigma_{A:B}}}$ as well.  Without loss of generality, let us consider the entanglement entropy of region $A$. Since the corresponding minimal surface of $S(A)$ is $\Gamma_A^{min}$, we just need to prove $\Gamma_A^{min} \in \boldsymbol{\mathit{\Sigma_{A:B}}}$. Firstly, based on  the definition of  $\Gamma_A^{min}$,  we know that 
\begin{align}
    (\partial \Gamma_A^{min})|_{\partial M} &= \partial A, \\
    (\partial \Gamma_A^{min})|_{Q} &= \partial Q_A, \\
    \quad \partial M_A&= \Gamma_A^{min} \cup  \left(A\cup Q_A\right).
\end{align} 
Secondly, employing ``wedge nesting" we know $M_A \subset M_{AB}$,  so let
$\widetilde{A}=    \left(A\cup Q_A\right)$,  then we get \begin{align}
    \partial  \Gamma_A^{min}&=\partial\widetilde{A},\\
    M_A \subset M_{AB}, \quad \partial M_A&= \Gamma_A^{min} \cup \widetilde{A}.
\end{align}

Two conditions in (\ref{es}) for the elements in $\boldsymbol{\mathit{\Sigma_{A:B}}}$ are satisfied. Therefore, $\Gamma_A^{min}\in\boldsymbol{\mathit{\Sigma_{A:B}}}$, and one must have $S(A) \ge E_W(A:B)$, following the definition in (\ref{es}). Or intuitively, we may state that $\Gamma_A^{min}$ is a cross section which divides the entanglement wedge $M_{AB}$ into two parts and separates $A$ and $B$, by definition its area must not be smaller than the minimal cross section of $M_{AB}$, which is juts defined as $E_W(A:B)$.
For the same reason one can get $S(B)\ge E_W(A:B)$, thus the above inequality is proved.

\subsection{$E_W(A:B) \ge \frac{1}{2} I(A:B)$}

This inequality states that $E_W(A:B)$  must not be smaller than one half of the holographic mutual information between $A$ and $B$.  Since the mutual information is constructed by the entanglement entropy, explicitly $I(A,B)=S(A)+S(B)-S(AB)$,  we may change the inequality into the form
\begin{equation}
2E_W(A:B)+S(AB) \ge S(A)+S(B).\label{i2}
\end{equation}

\begin{figure}
\centering 
\includegraphics[width=1\textwidth]{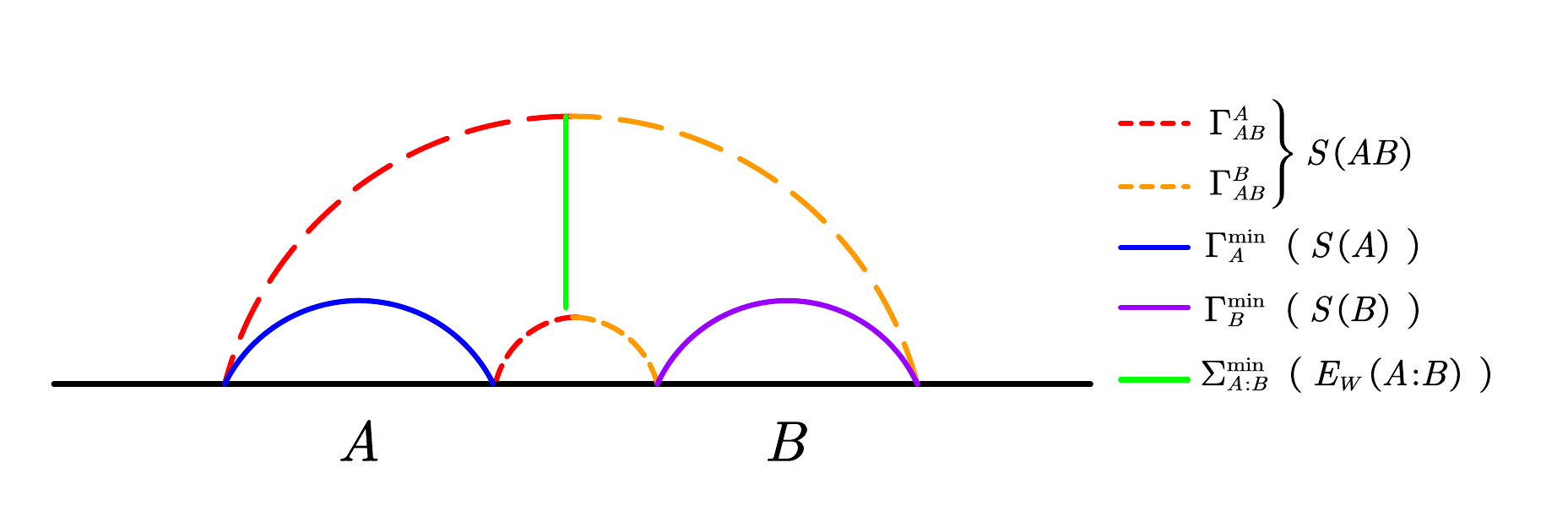}
\caption{\label{fig:ie2} 
The diagrammatic demonstration of inequality (\ref{i2}) in AdS/CFT. Since 
 the union of the green line and the red dotted line is homologous to $A$, while the blue line is the minimal surface homologous to $A$, we have $\mathcal{A}(\Gamma_{AB}^A)+E_W(A:B)\ge S(A)$. Similarly, 
 $\mathcal{A}(\Gamma_{AB}^B)+E_W(A:B)\ge S(B)$. Combining these two inequalities one can get $2E_W(A:B)+S(AB) \ge S(A)+S(B)$.   }
\includegraphics[width=1\textwidth]{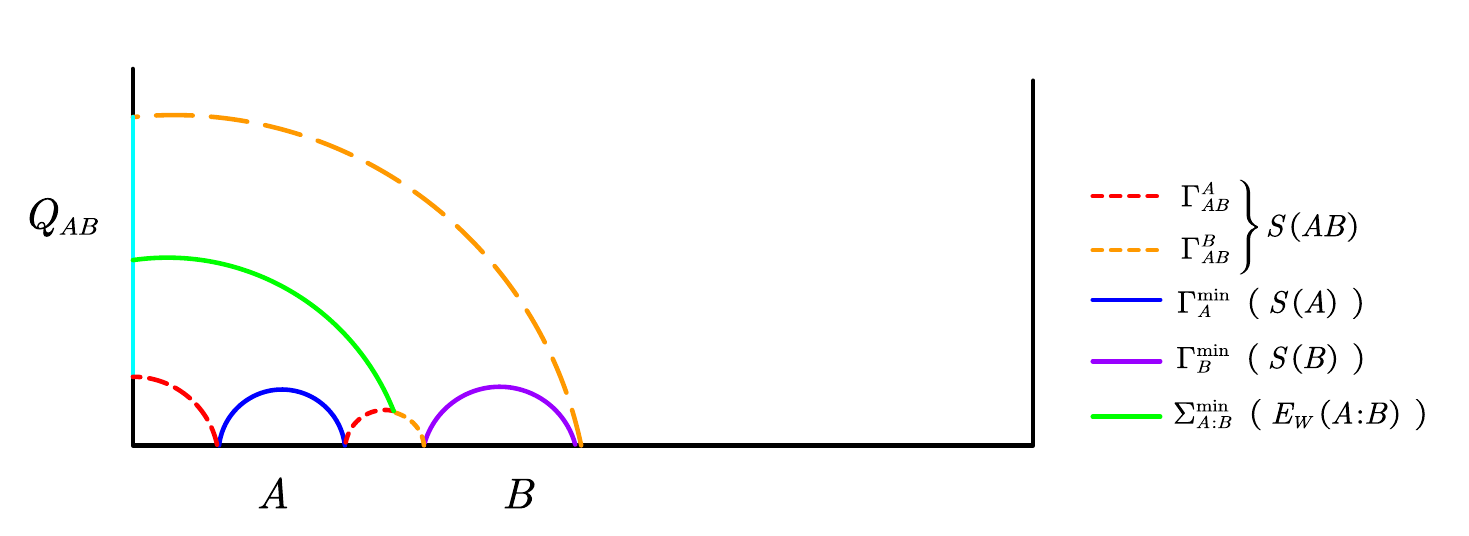}
\caption{\label{fig:ie2bcft}
The diagrammatic demonstration of inequality (\ref{i2}) in AdS/BCFT. }
\end{figure}

Two examples of  diagrammatic proof are demonstrated in Fig.(\ref{fig:ie2}) and (\ref{fig:ie2bcft}) for CFT and BCFT respectively, where one typical phase of $E_W(A:B)$ is plotted to demonstrate this inequality geometrically. Nevertheless, from Fig.(\ref{fig:bcftsabphases})  and Fig.(\ref{fig:bcftewcsdifferentphases}) one knows that there are various possible phases for $S(AB)$ and $E_W(A:B)$. If one insists to prove this inequality in a diagrammatic manner, then she/he has to enumerate all the possible phases case by case, that would be very cumbersome. Therefore, we  present the following proof in an algebraic manner.

By definition, the geometric correspondence of these quantities are given by
\begin{align}
 S(AB)=\mathcal{A}  \left(\Gamma_{AB}^{min}\right),\\ 
E_W(A:B)= \mathcal{A} (\Sigma_{A:B}^{min}).
 \end{align}
It is noticed that $\Sigma_{A:B}^{min}$ divides $\Gamma_{AB}^{min}$ into two parts $\Gamma_{AB}^{min}=\Gamma_{AB}^A \cup \Gamma_{AB}^B$.
Note the quantity on the left-hand side in (\ref{i2}) can be recombined  as 
\begin{equation}
 \mathcal{A} (\Gamma_{AB}^A+\Sigma_{A:B}^{min})+\mathcal{A} (\Gamma_{AB}^B+\Sigma_{A:B}^{min}).\end{equation}
 
We demonstrate this procedure in 
Fig.(\ref{fig:ie2}) for CFT case, while in Fig.(\ref{fig:ie2bcft}) for BCFT case. 
Now, our key observation is that $\left(\Gamma_{AB}^A+\Sigma_{A:B}^{min}\right)$ actually becomes a surface which belongs to the set $\boldsymbol{\mathit{\Gamma_A}}$, namely the collection of candidates for the minimal surface associated with subregion $A$ (See the example in Fig.(\ref{fig:ie2}) and (\ref{fig:ie2bcft})). That is,
 $\left(\Gamma_{AB}^A+\Sigma_{A:B}^{min}\right) \in \boldsymbol{\mathit{\Gamma_A}}$, since the conditions in the definition of $\boldsymbol{\mathit{\Gamma_A}}$  in (\ref{heeb}) are satisfied by $\left(\Gamma_{AB}^A+\Sigma_{A:B}^{min}\right)$. We check these conditions explicitly as follows. \par
For condition 1.  By the definition of $\Sigma _{AB}^{min}$, we have
 \begin{equation}
  \partial \Sigma _{AB}^{min} =\partial\widetilde{A}=\partial \left(A \cup \Gamma_{AB}^A\cup Q_{AB}^A\right).
\end{equation}
This is equivalent to\begin{align}
       \partial \left(\Sigma_{A:B}^{min} \cup {\Gamma }_{AB}^A\right)|_{\partial M}&=\partial A,\\
        \partial \left(\Sigma_{A:B}^{min} \cup {\Gamma }_{AB}^A\right)|_{Q}&=\partial Q_{AB}^A.
   \end{align}
So $\left(\Gamma_{AB}^A+\Sigma_{A:B}^{min}\right)$  satisfies condition 1 of $\boldsymbol{\mathit{\Gamma_A}}$ in (\ref{heeb}), and for the same reason we know $\left(\Gamma_{AB}^B+\Sigma_{A:B}^{min}\right)$  also satisfies condition 1 of $\boldsymbol{\mathit{\Gamma_B}}$.\par
   
For condition 2 in (\ref{heeb}). 
By the definition of $\Sigma_{A:B}^{min}$, there exists a corresponding $R_A \subset M_{AB}$, which satisfies 
    \begin{align}
    \partial R_A=\Sigma_{A:B}^{min}\cup\widetilde{A}.
    \end{align}
From above formulae we can get
\begin{equation} R_A \subset M  \ \ \ \ \ \text{and } \ \ \partial
    R_A=\left(\Sigma^{min}_{A:B}\cup\Gamma_{AB}^A\right) \cup A \cup Q_{AB}^A.\end{equation}
    
Therefore, condition 2 in (\ref{heeb}) is also satisfied. We conclude that $(\Gamma_{AB}^A+\Sigma_{A:B}^{min})$ belongs to $\boldsymbol{\mathit{\Gamma_A}}$ indeed. As a result, we must have
\begin{equation}\mathcal{A} (\Gamma_{AB}^A+\Sigma_{A:B}^{min}) \ge \mathcal{A} (\Gamma_A^{min}). \end{equation}

Similarly, we also have
\begin{equation} \mathcal{A} (\Gamma_{AB}^B+\Sigma_{A:B}^{min}) \ge \mathcal{A} (\Gamma_B^{min}). \end{equation}
Finally, combining these two inequalities we can prove the initial inequality on the relation between $E_W$  and the holographic mutual information 
\begin{equation}\mathcal{A} (\Gamma_{AB}^A+\Sigma_{A:B}^{min})+\mathcal{A} (\Gamma_{AB}^B+\Sigma_{A:B}^{min}) =2E_W(A:B)+S(AB) \ge  S(A)+S(B).\end{equation}

\subsection{$E_W(A:BC) \ge  E_W(A:B)$} \label{3}
\begin{figure}
\centering 
\includegraphics[width=1\textwidth]{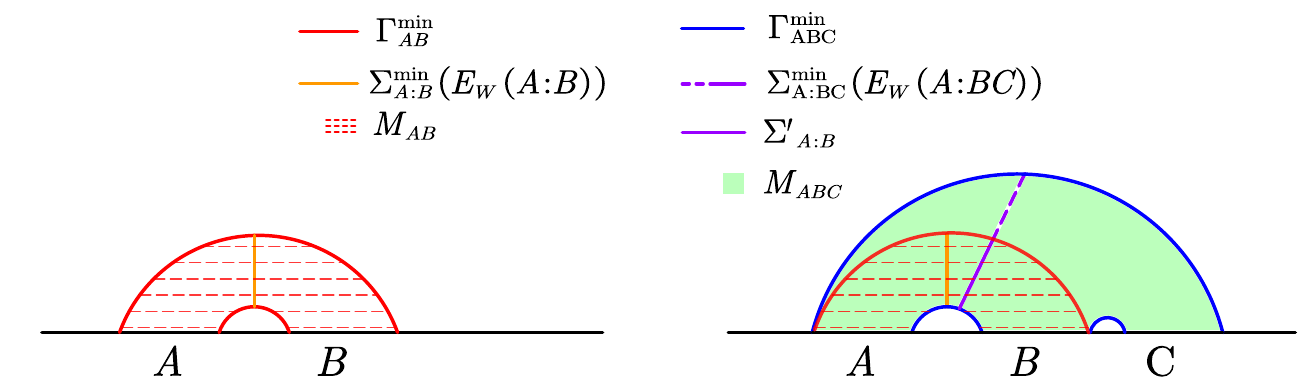}
\caption{\label{fig:ie3}
A diagrammatic demonstration of the proof for inequality $E_W(A:B)\le E_W(A:BC)$, which is geometrically described by the fact that the area of the orange line ($\Sigma^{min}_{A:B}$) is always less than that of the purple solid line ($\Sigma'_{A:B}$), which is just a part of $\Sigma^{min}_{A:BC}$, as illustrated in the right plot. Here for the clearness a simple configuration is plotted, while it should be stressed that the procedures of our proof are quite general and applicable for other configurations which may be very complicated.} 

\includegraphics[width=1\textwidth]{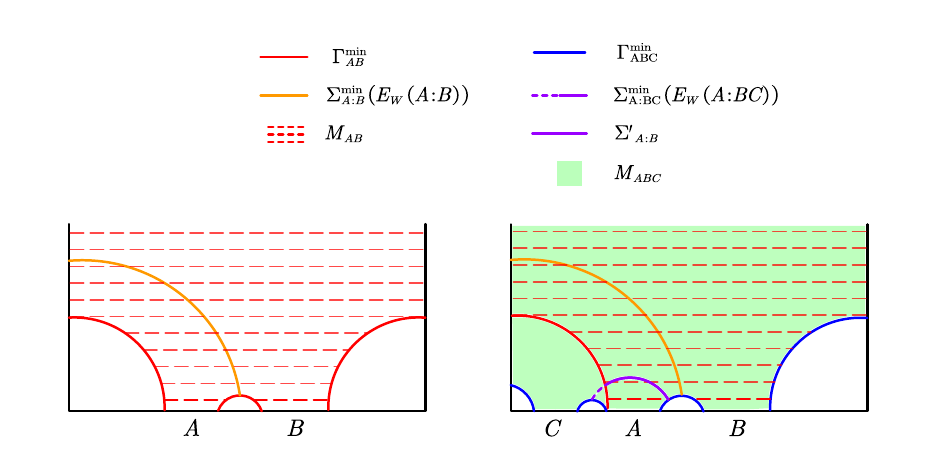}
\caption{\label{fig:ie3bcft} 
A diagrammatic demonstration of the proof for inequality $ E_W(A:BC)\ge E_W(A:B)$ in AdS/BCFT, which is geometrically described by the fact that the area of the orange line ($\Sigma^{min}_{A:B}$) is always less than that of the purple line ($\Sigma'_{A:B}$). Notice that there is a phase transition for $E_W$ from $\Sigma^{min}_{A:B}$ (which ends on the $Q$ brane) to $\Sigma^{min}_{A:BC}$ (which ends on the minimal surface $\Gamma^{min}_{ABC}$).
}\end{figure}

This inequality means that enlarging any one of region A and B does not decrease the correlations between these two regions. This inequality can be intuitively understood as follows. The minimal cross section of $E_W(A:BC)$ separates the entanglement wedge of ABC, namely $M_{ABC }$, into two parts $R_A$ and $R_{BC}$ which include A and BC, respectively.  At the same time it must separate $M_{AB}$ into  $R'_A$ and $R'_{B}$ as well, as illustrated in Fig.(\ref{fig:ie3}) and Fig.(\ref{fig:ie3bcft}), but $E_W(A:B)$ is the minimal area to separate $M_{AB }$ into any possible $R'_A$ and $R'_{B}$, so we should have $E_W(A:BC) \ge  E_W(A:B)$.

 To prove the above inequality in an algebraic method, our strategy is to show that one portion of the left-hand side of the inequality corresponds to the area of surface $\Sigma'_{A:B}$ which belongs to $\boldsymbol{\mathit{\Sigma_{A:B}}}$, while the right-hand side is the area of the minimal surface in $\boldsymbol{\mathit{\Sigma_{A:B}}}$.  Then one obtains the relations between their areas as $E_W(A:BC) \ge \mathcal{A}(\Sigma'_{A:B}) \ge E_W(A:B)$. (See  Fig.(\ref{fig:ie3})(\ref{fig:ie3bcft}).) \par

Now we need to construct such $\Sigma'_{A:B}$. In fact,  $\Sigma'_{A:B}$ may be defined as ``cut $\Sigma_{A:BC}^{min}$'', which is the part of $\Sigma_{A:BC}^{min}$ in the interior of $M_{AB}$, and given by \begin{equation}
  \Sigma'_{A:B}=M_{AB} \cap \Sigma_{A:BC}^{min}. \label{ cut EW} 
\end{equation}
Next we just need to show that $  \Sigma'_{A:B}$ belongs to $\boldsymbol{\mathit{\Sigma_{A:B}}}$.
For this purpose we just check that $\Sigma'_{A:B}$ satisfies two conditions in  the definition of $\boldsymbol{\mathit{\Sigma_{A:B}}}$. \par

Firstly, by the definition of $\Sigma_{A:BC}^{min}$, there exists a bulk region $R_A$ which satisfies 
\begin{align}
R_A\subset M_{ABC},\\
 \partial R_A=\widetilde{A}\cup\Sigma_{A:BC}^{min}. 
\end{align}
Then we define  ``cut $R_A$'' to be the part of $R_A$ in $M_{AB}$ as

\begin{equation}
R'_A= R_A \cap M_{AB}.
\end{equation}

\begin{figure}
\centering 
\includegraphics[width=1\textwidth]{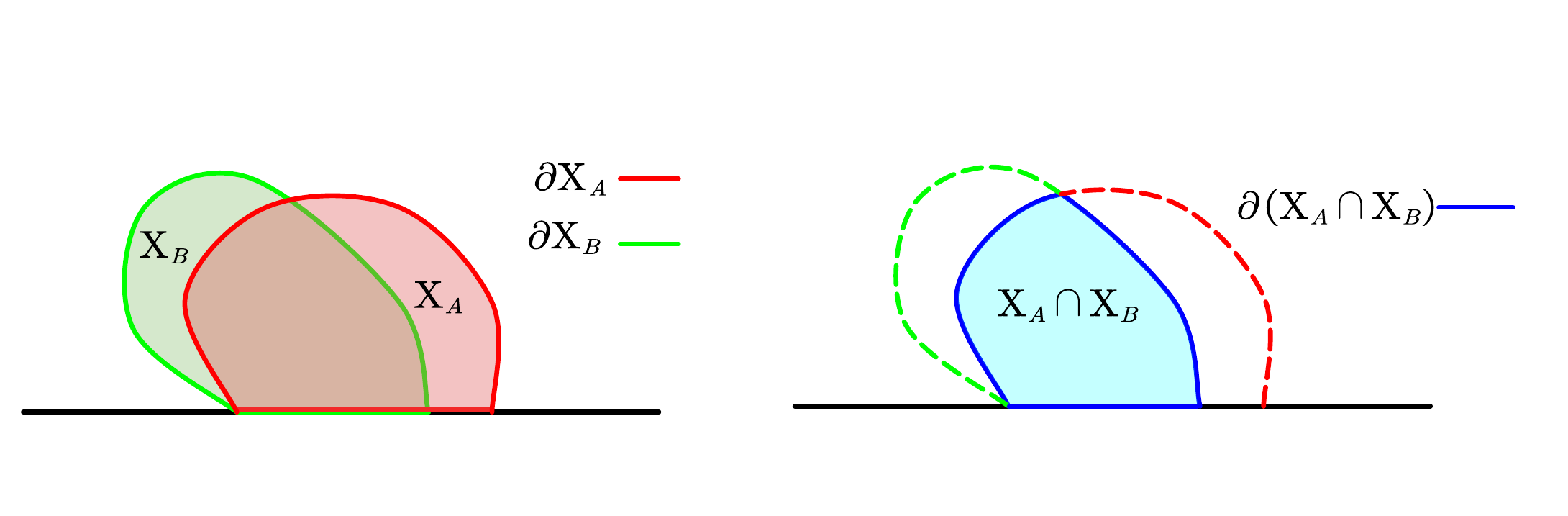}
\caption{\label{fig:UaL} 
The boundary of $X_A\cap X_B$ consists of three parts: one part is the boundary of $X_A$ in the interior of $X_B$, and  the second part is the boundary of  $X_B$ in the interior of $X_A$, while the third part is the common boundary of $X_A$ and $R_B$.}
\end{figure}

Now we intend to argue that for $\Sigma'_{A:B}$, $R'_A$ is exactly the bulk region such that one has $\partial R'_A=  \Sigma'_{A:B} \cup \widetilde{A'} $. (Notice that here $ \widetilde{A'} $ is different from the above $ \widetilde{A}$. The former is associated with $\Sigma_{A:B}^{min}$, while the latter is associated with $\Sigma_{A:BC}^{min}$.) The existence of such a $R'_A$ will lead to $\Sigma'_{A:B}\in\boldsymbol{\mathit{\Sigma_{A:B}}}$. To prove this explicitly, we consider the boundary of ``cut $R_A$'' 
\begin{equation}
\partial R'_A=\partial \left(M_{AB} \cap R_A\right).
\end{equation}
Now we intend to figure out what one would obtain on the right-hand side of this equation. It is noticed that in general the boundary of the intersection of two regions  $X_A$ and $X_B$ is the union of the boundary of $X_A$ in the interior of $X_B$, the boundary of $X_B$ in the interior of $X_A$ and the common boundary of two regions $X_A$ and $X_B$, as illustrated in Fig.(\ref{fig:UaL}).   
Similarly, in our case we find the boundary of $R'_A$ contains the following three parts. Firstly, from the wedge nesting property $M_{AB} \subset M_{ABC}$ and the definition of $R_A$, one can get the boundary of $R_A$ in the interior of $M_{AB}$ is just  $\Sigma'_{A:B}$; secondly, according to the definition of $\Gamma^{A}_{AB}$, the boundary of $M_{AB}$ in the interior $R_A$ must be one choice of  $\Gamma^{A}_{AB}$; thirdly, it is obviously seen that region $A$ is the unique common boundary of $M_{AB}$ and $R_A$. As a result, we have the following expressions
\begin{align}
\partial R'_A&=\partial \left(M_{AB} \cap R_A\right)\\&=  \Sigma'_{A:B} \cup \Gamma_{AB}^A \cup A\\&=  \Sigma'_{A:B} \cup \widetilde{A'}.\
\end{align}
From the above results, we know there exists a bulk region $R'_A \subset M_{AB}$, such that $\partial R'_A=  \Sigma'_{A:B} \cup \widetilde{A'} $. In addition, since $R'_A$ is a closed region, we have  $\partial  \Sigma'_{A:B}=\partial   \widetilde{A'}$. Therefore, we conclude that $ \Sigma'_{A:B}$  belongs to $\boldsymbol{\mathit{\Sigma_{A:B}}}$, then the inequality in this subsection so far is proved.

\subsection{$E_W(A:BC)\ge \frac{1}{2}  I(A:B)+\frac{1}{2}I(A:C)$}

 Having proved the former inequalities, we find the proof of this inequality becomes pretty easy. Firstly, we just apply the known inequality (\ref{inequality 2})
\begin{equation}
E_W(A:B) \ge \frac{1}{2}I(A:B).
\end{equation}
Replacing $B$ with $BC$ in this inequality, we get 
\begin{equation}
E_W(A:BC) \ge \frac{1}{2}I(A:BC).
\end{equation}
Then with the use of the holographic  MMI inequality\cite{Hayden:2011ag}\cite{Chou:2020ffc} \begin{equation}
I(A:BC)\ge I(A:B)+I(A:C),
\end{equation}

we prove this inequality \begin{equation}
E_W(A:BC)\ge \frac{1}{2}  I(A:B)+\frac{1}{2}I(A:C).
\end{equation}

\subsection{$E_W(AB:CD)\ge E_W(A:C)+E_W(B:D)$}

In this inequality we notice that two terms on the right-hand side are nothing but the areas of minimal-area elements in $\boldsymbol{\mathit{\Sigma_{A:C}}}$ and $\boldsymbol{\mathit{\Sigma_{B:D}}}$, namely $\mathcal{A}(\Sigma_{A:C}^{min})$ and $\mathcal{A}(\Sigma_{B:D}^{min})$, while the term on the left-hand side is just  $\mathcal{A}(\Sigma_{AB:CD}^{min})$. So 
 our strategy is to prove that there exist two non-overlapping surfaces $X$ and $Y$ contained in  $\Sigma_{AB:CD}^{min}$. Moreover, $X$ and $Y$  belong to $\boldsymbol{\mathit{\Sigma_{A:C}}}$ and $\boldsymbol{\mathit{\Sigma_{B:D}}}$, respectively. Then we can get the following relations \begin{align}
        E_W(AB:CD)= \mathcal{A}(\Sigma_{AB:CD}^{min})\ge \mathcal{A}(X)+\mathcal{A}(Y)\ge E_W(A:C)+E_W(B:D).
    \end{align}         
So next our target is to show the existence of such $X$ and $Y$ contained in $\Sigma_{AB:CD}^{min}$.\par
    
Firstly, exploiting ``wedge nesting" we can get
\begin{equation}
AC \cap BD= \emptyset \Rightarrow  M_{AC} \cap M_{BD}=\emptyset.
\end{equation}
As in (\ref{ cut EW}), we may define two ``cut $\Sigma_{AB:CD}^{min}$''
\begin{align}
\Sigma'_{A:C}=\Sigma^{min}_{AB:CD} \cap M_{AC},\\
\Sigma'_{B:D}=\Sigma^{min}_{AB:CD} \cap M_{BD}.
\end{align}
Because
\begin{equation}
M_{AC} \cap M_{BD}=\emptyset,
\end{equation}
we have\begin{equation}
\Sigma'_{A:C}\cap \Sigma'_{B:D}=\emptyset.
\end{equation}
It actually means that \begin{equation}
\Sigma^{min}_{AB:CD}=\Sigma'_{A:C} \cup \Sigma'_{B:D} \cup (\text{the remaining surfaces}).
\end{equation}
Next we need to prove\begin{align}
\Sigma'_{A:C}\in \boldsymbol{\mathit{\Sigma_{A:C}}},\\
\Sigma'_{B:D}\in \boldsymbol{\mathit{\Sigma_{B:D}}}.
\end{align}
Because $M_{AC}$ does not overlap with $M_{BD}$, one can prove above two formulae independently. Moreover, based on the definition of $\Sigma'_{A:C}$ and $\Sigma'_{B:D}$, one can prove each of them belongs to set $\boldsymbol{\mathit{\Sigma_{A:C}}}$ or $\boldsymbol{\mathit{\Sigma_{B:D}}}$ in a parallel way as we performed in subsection (\ref{3}). Then finally we have \begin{equation}
E_W(AB:CD)\ge \mathcal{A} (\Sigma'_{A:C})+\mathcal{A} (\Sigma'_{B:D})\ge E_W(A:C)+E_W(B:D).\label{ie5}
\end{equation}\par
We remark that this inequality is just satisfied by quantum states with holographic dual, not applicable to  arbitrary quantum states\cite{Bao:2018gck}. So it may be used to justify whether a quantum state in CFT has a smooth bulk dual.  

\section{Conclusion and Discussion}
In this paper we have investigated the inequalities for the entanglement wedge cross section in the context of AdS/BCFT duality. We have constructed the candidate sets for HEE and $E_W$ in general cases, and developed an algebraic method to show that $E_W$ satisfies the same inequalities as in AdS/CFT. We have only considered the case that the bulk is pure AdS vacuum. As a matter of fact, one can show that in AdS black hole background (which is dual to BCFT at finite temperature) these inequalities should still hold. It is quite interesting to compare the algebraic method and the diagrammatic method in the proof of these inequalities. The diagrammatic method is applied naturally in the geometric description of entanglement entropy, and the relevant inequalities can be understood and then proved intuitively in this way. However, when the system consists of multi partitions and the subregion is the union of many disconnected parts, then the corresponding diagram would become complicated. In particular, when the CFT has a boundary as in the context of AdS/BCFT, the entanglement exhibits abundant phase structure, and the diagrammatic method suffers from the enumeration of all the possible diagrams involved in inequalities. In this situation, the algebraic method is very suitable for proving these inequalities because the process of proof does not depend on the specific phases of the entanglement. The price that one has to pay is that the proof might become abstract. Therefore, in this paper we have also illustrated some typical diagrams to assist us to understand the process of proof.   

Our work in this paper may be pushed forward in following directions in future. 
Firstly, one may check the inequalities of $E_W$ in time-dependent bulk gravity, which  could improve our  understanding on the dynamical properties of $E_W$. Secondly, it is very desirable to apply the algebraic method proposed in this work to explore more inequalities of $E_W$, especially holographic $E_W$ inequalities such as (\ref{ie5}), which might impose new constraints on holographic states.
Thirdly, it is interesting to consider the $E_W$ inequalities in other gravity theories where the minimization prescriptions for the entanglement entropy is modified. Finally, it is also very interesting to consider the inequalities beyond the leading order of the holographic entanglement. For instance, one may take 
into account the bulk corrections to the holographic entanglement as discussed in  \cite{Akers:2019lzs}\cite{Akers:2021lms}, and then check whether the behaviours of those inequalities  would be changed.

\section*{Acknowledgments}

We are very grateful to Yuxuan Liu and Zhuo-Yu Xian for helpful discussions.
This work is supported in part by the Natural Science Foundation
of China under Grant No.~11875053 and 12035016. It is also supported by Beijing Natural Science Foundation under Grant No. 1222031.

\end{document}